\documentstyle[pre,aps,preprint]{revtex}
\begin{document}
\draft
\title{Motion of a driven tracer particle in a one-dimensional
symmetric lattice gas.}

\author{S.F. Burlatsky}
\address{Department of Chemistry 351700,
 University of Washington,
 Seattle, WA 98195-1700 USA}
\author{G. Oshanin and M. Morea}
\address{Laboratoire de Physique Th\'eorique des Liquides,
Universit\'e Paris VI, 4 Place Jussieu, 75252 Paris Cedex 05,
France}
\author{W.P. Reinhardt}
\address{Department of Chemistry 351700,
 University of Washington,
 Seattle, WA 98195-1700 USA}
\maketitle
\begin{abstract}
Dynamics of a tracer particle subject to a constant driving
force
$E$   in a one-dimensional
lattice gas of hard-core particles whose transition rates are symmetric.
We show that the mean displacement of the driven
tracer, $\overline{X_{T}(E,t)}$, grows in time $t$
as $\overline{X_{T}(E,t)} = \sqrt{\alpha t}$,  rather than the linear time
dependence found for non-interacting  (ghost)  bath particles.
The prefactor $\alpha$ is determined implicitly,  as the solution of a
transcendental equation,
for an arbitrary magnitude of the driving force and an arbitrary
concentration of the lattice gas particles. In  limiting cases
the prefactor is  obtained explicitly. Analytical predictions are seen to
be in a good
agreement with the results of numerical simulations.
\end{abstract}
\section{Introduction}

Dynamic and equilibrium properties of lattice gases, i.e. systems involving
randomly moving particles with hard-core interactions, have received much
interest within the last several decades. A number of important theoretical
results have been obtained for such systems revealing non-trivial, many-body
behavior \cite{Ligget,Katz,fer,bei,Haus,spohn,jan,dera,derc,Blumen,per,kehr}%
. Lattice gas models often serve as microscopic models of complex
physical phenomena. To name a few we mention dynamics of motor proteins \cite
{prost,dere}, growth of interfaces \cite{krug,maj,kand}, traffic jams and
queuing problems \cite{kip}. A lattice-gas approach has been used for the
derivation of Euler-type hydrodynamic equations, e.g. the Burgers equation
\cite{leba,beic}. Another important example concerns the spreading of
molecularly thin wetting films, where experimental studies
\cite{hesa,hesb,hesc}
have evidenced surprising universal laws which recently have been explained
in terms of a lattice gas model \cite{bura,burb,burc}.
We believe also that a robust microscopic, molecular approach to such
physical phenomena as shear-induced ordering in colloidal suspensions \cite
{pusey} or stick-slip motion of mica planes separated by an ultrathin liquid
layer \cite{grana,granb,isra,robb} could also begin with a description based
on a lattice-gas picture.

At present time two models are well-studied in the literature. In the first,
the so-called asymmetric exclusion process, all particles in the system
perform stochastic motion, constrained by hard-core interactions, in the
field of a constant driving force \cite
{Ligget,Katz,fer,bei,Haus,spohn,jan,dera,derc,masi,kuta,derb}. Here the
velocity, diffusion constant and equilibrium configurations have been
calculated exactly for different types of boundary conditions ( e.g. \cite
{Katz,jan,dera,derb} and references therein). In the second, no external
force is present and all particles have symmetric transition rates.
Remarkably, in such a situation, the motion of a labeled, tracer particle is
non-diffusive in low dimensions. For example, the mean-square displacement %
$\overline{X^{2}_{T}(E = 0,t)}$ of a tracer particle (identical except for
its observability to all other particles) in a one-dimensional (1D)
symmetric lattice gas shows a sublinear growth with time \cite{Harris,alex}
\[
\overline{X^{2}_{T}(E = 0,t)} \; = \; \frac{C_{0}}{1 - C_{0}} \; (\frac{2 t}{%
\pi})^{1/2}, \quad (1)
\]
where $C_{0}$, $0\leq C_0<1$, denotes the mean (constant) concentration of
vacant sites and the argument $E = 0$ signifies that the external force is
absent and that all  particles have symmetric transition rates. Hence, in 1D
trajectories $X_{T}(E = 0,t)$ of  such a tracer particle are more compact than
these of particles without the hard-core constraints. In 2D the mean-square
displacement $\overline{X^{2}_{T}(E = 0,t)}$ shows a linear dependence on
time with additional logarithmic terms \cite{per,henk}; in 3D it grows
linearly in time with the diffusion constant being a nontrivial function of
the particle concentration.

In the present paper we focus on the less studied and less understood
situation in which only one particle (the tracer) experiences the action of
an external (constant) driving force, $E$ and thus has asymmetric transition
rates, while all other particles (bath particles) are not subject to this
force and have symmetric transition rates. The tracer behavior in such a
system with a vanishingly small driving force has been first examined in \cite
{fer} in which the question of the validity of the Einstein relation for the
hard-core lattice gases has been addressed\footnote{%
This point will be discussed in more detail in Section IV.D.}. Such a model
has been used  \cite{AlexanderF} in a numerical study  of
the gravity driven motion of a finite rigid rod in a ''sea'' of hard-core
monomers.  Another physical example corresponds
to the situation in which a charged particle diffuses in a lattice gas of
electrically neutral particles in the presence of a constant electric field %
$E$. The extreme case of infinitely strong electric fields ($E=\infty $),
which means that the tracer particle may move only in one direction, has
been studied in \cite{burd}. It has been shown, for example, that in 1D
systems the mean displacement, $\overline{X_T(E=\infty ,t)}$, of a charged
tracer particle grows sub linearly with time,

\[
\overline{X_T(E=\infty ,t)}\;\propto \;(\alpha _\infty \;t)^{1/2},\quad (2)
\]
where $\alpha _\infty $ is a constant \cite{burd}. Eq.(2) indicates that
hard-core interactions give rise to an effective friction. In the
one-dimensional case this force is much stronger than the viscous friction
for one particle and grows in proportion to the mean displacement of the
tracer , i.e. as $\sqrt{t}$,  as this is a measure of the size of the
compressed region preceding
the tracer which hinders the ballistic motion of the driven
particle.

Here we study motion of a driven tracer particle in a symmetric lattice gas
in the general case of fields of arbitrary strength and at arbitrary
concentrations of the lattice-gas particles. Focusing on one-dimensional
situations in which the hindering effect of the lattice-gas (bath) particles
on the tracer motion is most pronounced, we devise a mean-field-type theory
which allows simple calculation of the mean displacement of the tracer
particle as a function of time and other pertinent parameters. We find that $%
\overline{X_T(E,t)}$ has the following dependence

\[
\overline{X_T(E,t)}\;=\;(\alpha \;t)^{1/2},\quad (3)
\]
where the parameter $\alpha $ is a time-independent constant, which is a
complicated function of field strength $E,$ which determines the transition
probabilities $p$ and $q$, and concentration of the bath particles, $C_p,$.
This constant is determined here for arbitrary values of $E$ and
$C_p$. Our analytical findings are in excellent agreement with the results
of numerical simulations.

The paper is structured as follows: In Section 2 we describe the model. In
Section 3 we present definitions and write down basic equations describing
motion of particles. In Section 4 we determine explicitly the growth law for
the mean displacement of the tracer particle, Eq.(3), and evaluate a closed
transcendental equation for the parameter $\alpha $. In several limiting
cases the dependence of $\alpha $ on the pertinent parameters is explicitly
obtained. Section 5 presents results of numerical simulations and comparison
of these with our analytical predictions. Finally, In Section 6 we conclude
with summary of our results and discussion.

\section{The model.}

The model is defined in the following way. Consider a one-dimensional
regular lattice of unit spacing, infinite in both directions, the sites $%
\{X\}$ of which are either singly occupied by identical particles or
vacant. The particles
are initially placed at random (constrained by the condition that double
occupancy of sites is forbidden) with mean concentration $C_p=1-C_0,$ where $%
C_p$ is the mean site occupancy, $C_0$ being the mean site vacancy. The
tracer particle is put initially at the origin, i.e. at $X=0$. A
configuration of the system is characterized by an infinite set of
(time-dependent) occupation variables $\{\tau _X\}$, where $\tau _X=1$ if
site $X$ is occupied and $\tau _X=0$ if site $X$ is vacant. Consequently,
the variable $\eta _X=1-\tau _X$ describes the probability that site $X$ is
vacant.

The dynamics of the bath particles is symmetric: each particle waits a
(random) exponentially distributed time with mean $1$ and then attempts to
jump, with equal probability $(1/2)$ to the right or left neighboring site.
The jump actually occurs if the chosen site is empty. The tracer particle
motion
is asymmetric: the tracer waits a random exponentially distributed time with
mean $1$ and then randomly selects a jump direction. It chooses right-hand
(left-hand) adjacent site with probability $p$ ($q=1-p$). The jump occurs if
the selected site is vacant. If the asymmetry in tracer jump probabilities
is due to an external electric field, $E$, one has the relation $p/q=\exp
(\beta E)$, where $\beta $ is the inverse temperature. For simplicity we
have set the tracer charge to unity. We also assume, without lack of
generality, that $E$ is oriented in the positive direction, i.e. $E>0$, and
thus $p>q$.

\section{Definitions and basic equations.}

We start by writing the equations which describe dynamics of the tracer
particle. Let $X_T(E,t)$ denote the position of the tracer at time $t$ (by
definition $X_T(0)=0$) and $P(X,t)$ be the probability that the tracer is at
site $X$ at time $t$. The mean displacement of the tracer particle, i.e. $%
\overline{X_T(E,t)}$, is then defined as

\[
\overline{X_{T}(E)} \; = \; \sum_{X} X \; P(X,t) \quad (4)
\]

The time evolution of $P(X,t)$ is governed by the equation

\[
\dot{P}(X,t) \; = \; - \; P(X,t) \; (p \; \eta_{X+1}(t) \; + \; q \;
\eta_{X-1}(t)) \; + \quad
\]
\[
+ \; \eta_{X}(t) \; (p \; P(X-1,t) \; + \; q \; P(X+1,t)), \quad (4)
\]
where the dot denotes the time derivative. The first two terms on
the r-h-s of Eq.(4) describe respectively the change in $P(X,t)$ due to
jumps of the tracer particle from the site $X$ to sites $X \pm 1$, while
the third and fourth
terms  account for jumps from the sites $X \pm 1$ to
the site $X$. Multiplying both sides of Eq.(4) by $X$ and summing over all
lattice sites we arrive at  the rate equation

\[
\overline{\dot{X}_{T}(E,t)} \; = \; p f_{1} \; - \; q f_{-1}, \quad (5)
\]
where

\[
f_\lambda \;=\;\sum_XP(X,t)\;\eta _{X+\lambda }(t)\quad (6)
\]
is the pairwise tracer-vacancy correlation function, which can be thought
off as the probability of finding at time $t$ a vacancy at the distance $%
\lambda $ from the tracer. The correlation function is defined in the frame
of reference moving with the tracer and jumps of the tracer change the value
of $f_\lambda $.

Consider now evolution of $f_\lambda $, which completely determines the mean
displacement of the tracer, Eq.(5). The change in $f_\lambda $ results from
two different processes

\[
\dot{f}_\lambda \;=\;\hat{L}_{bath}(f_\lambda )\;+\;\hat{L}_{trac}(f_\lambda
),\quad (7)
\]
where the operator $\hat{L}_{bath}$ accounts for the contribution coming
from the motion of bath particles, whilst $\hat{L}_{trac}$ describes the
contribution of the tracer motion itself. Explicitly, for $\hat{L}%
_{bath}(f_\lambda )$ we have (for $|\lambda |>1$)

\[
\hat{L}_{bath}(f_\lambda )\;=\;\frac 12(1\;-\;f_\lambda )(f_{\lambda
-1}\;+\;f_{\lambda +1})\;-\;\frac 12f_\lambda (1\;-\;f_{\lambda
-1}\;+\;1\;-\;f_{\lambda +1}),\quad (8)
\]
where the first term describes the ''birth'' of a vacancy at the occupied
site $X+\lambda $ due to the jumps of a bath particle from $X+\lambda $ to
vacant sites $X+\lambda \pm 1$; the second term describes a ''death'' of a
vacancy at a vacant site $X+\lambda $ due to the jumps of a bath particle
from sites $X+\lambda \pm 1$. One readily notices that the nonlinear terms
in Eq.(8) cancel each other and $\hat{L}_{bath}$ is simply the second finite
difference operator (for $|\lambda |>1$)

\[
\hat{L}_{bath}(f_\lambda )\;=\;\frac 12(-2f_\lambda \;+\;f_{\lambda
-1}\;+\;f_{\lambda +1})\quad (9)
\]
The ''diffusive'' free particle like behavior in Eq.(9) is, of course, the
consequence of the fact that all bath particles are identical, leading to
cancellation of non - linear terms in Eq.(8).

Consider now the contribution due to the motion of the tracer particle. In
an explicit form we have (for $|\lambda| > 1$)

\[
\hat{L}_{trac}(f_\lambda )\;=\;q\;f_{-1}\;(1-f_\lambda )\;f_{\lambda
-1}\;-\;q\;f_{-1}\;f_\lambda \;(1-f_{\lambda -1})\;+\quad
\]
\[
\;+\;p\;f_1\;(1-f_\lambda )\;f_{\lambda +1}\;-\;p\;f_1\;(1-f_{\lambda
+1})\;f_\lambda ,\quad (10)
\]
where the terms on the r-h-s of Eq.(10) describe respectively the following
events: (a) an $occupied$ site at the distance $\lambda $ from the tracer
becomes $vacant$ if the tracer jumps into the previously $vacant$ left-hand
adjacent site and the site at distance $\lambda -1$ is $vacant$; (b) a $%
vacant$ site at distance $\lambda $ from the tracer becomes $occupied$, if
the tracer jumps into the previously $vacant$ left-hand adjacent site and
the site at distance $\lambda -1$ is $occupied$; (c) an $occupied$ site at
distance $\lambda $ becomes $vacant$ when the tracer jumps into the
previously $vacant$ right-hand site and the site at distance $\lambda +1$ is
$vacant$; eventually, (d) a $vacant$ site at distance $\lambda $ becomes $%
occupied$ when the tracer jumps into the previously $vacant$ right-hand site
and the site at distance $\lambda +1$ is $occupied$.

Next, similar reasoning gives the behavior of $f_\lambda $ in the immediate
neighborhood of tracer, i.e. at sites with $\lambda =\pm 1$, which can
be thought off as the boundary conditions for Eq.(7). Time evolution of $%
f_{\pm 1}$ again is due to both the motion of the bath particles and of the
tracer, i.e. can be represented in the form of Eq.(7), but here the
operators $\hat{L}_{bath}$ and $\hat{L}_{trac}$ are given as

\[
\hat{L}_{bath}(f_{\pm 1})\;=\;\frac 12\;f_{\pm 2}\;(1-f_{\pm 1})\;-\;\frac 12%
\;f_{\pm 1}\;(1-f_{\pm 2}),\quad (11)
\]
\[
\hat{L}_{trac}(f_1)\;=\;-\;p\;f_1\;(1-f_2)\;+\;q\;f_{-1}\;(1-f_1)\quad (12)
\]
and
\[
\hat{L}_{trac}(f_{-1})\;=\;p\;f_1\;(1-f_{-1})\;-\;q\;f_{-1}\;(1-f_{-2})\quad
(13)
\]
Explicitly, we have for $f_{\pm 1}$ the following equations

\[
\dot{f}_{1} \; = \; \frac{1}{2} \; f_{2} \; (1 - f_{1}) \; - \; \frac{1}{2}
\; f_{1} \; (1 - f_{2}) \; - \; p \; f_{1} \; (1 - f_{2}) \; + \; q \;
f_{-1} \; (1 - f_{1}) \quad (14)
\]
\[
\dot{f}_{-1} \; = \; \frac{1}{2} \; f_{-2} \; (1 - f_{-1}) \; - \; \frac{1}{2%
} \; f_{-1} \; (1 - f_{-2}) \; + \; p \; f_{1} \; (1 - f_{-1}) \; - \; q \;
f_{-1} \; (1 - f_{-2}) \quad (15)
\]

Another pair of boundary conditions for Eq.(10) can be obtained by assuming
that correlations between $P(X,t)$ and $\eta _{X+\lambda }(t)$ vanish in the
limit $\lambda \rightarrow \pm \infty $, which seems quite plausible on
physical grounds, and which may be checked numerically. We have from Eq.(6)

\[
f_{\lambda \rightarrow \pm \infty }\;=\;\eta _{\pm \infty
}(t)\;\sum_XP(X,t)\;=\;C_0,\quad (16)
\]
where we have used the normalization condition $\sum_XP(X,t)=1$ and also the
assumption that at infinitely large separations from the tracer the mean
concentration of vacancies is unperturbed and equal to its equilibrium value
$C_0$.

Eqs.(5), (7) to (13) constitute a closed system of equations for the
derivation of $\overline{X_T(E,t)}$, within the framework outlined.
Solutions to these equations will be discussed in the next section.

\section{Solution of dynamic equations and analytical results.}

We first turn to the continuous-space description and rewrite our equations
in the limit of continuous-$\lambda$. Then, for $\lambda \neq \pm 1$ we have
for the operator

\[
\hat{L}_{bath}(f_\lambda )\;\approx \;\frac 12\;\frac{\partial ^2f_\lambda }{%
\partial \lambda ^2},\quad (17)
\]
and, at points $\lambda =\pm 1$, Eq.(11) yields

\[
\hat{L}_{bath}(f_{\pm 1})\;\approx \left. \;\frac 12\;\frac{\partial
f_\lambda }{\partial \lambda }\right| _{\lambda =\pm 1},\quad (18)
\]

Next, expanding $f_{\lambda-1} \approx f_{\lambda} - \partial
f_{\lambda}/\partial \lambda$ and $f_{\lambda+1} \approx f_{\lambda} +
\partial f_{\lambda}/\partial \lambda$ we obtain for the operator $\hat{L}%
_{trac}$, Eq.(10), the following approximate expression
\[
\hat{L}_{trac}(f_\lambda )\;\approx \;-\;q\;f_{-1}\;\frac{\partial f_\lambda
}{\partial \lambda }\;+\;p\;f_1\;\frac{\partial f_\lambda }{\partial \lambda
}\;=\;\overline{\dot{X}_T(E,t)}\;\frac{\partial f_\lambda }{\partial \lambda
},\quad (19)
\]
which holds for $\lambda \neq \pm 1$.

The approximation as in Eq.(19), in which the coefficient before the
gradient term is equal to the $mean$ velocity of the tracer particle, is
equivalent to the neglect of fluctuations in the tracer trajectory $X_T(E,t)$%
. In the following we show, however, that such a ''deterministic''
approximation is quite appropriate and leads to results which are in good
agreement with simulation data. The reason this approximation works here is
that in such a system the fluctuations of the tracer trajectories are
essentially suppressed. This is due to the accumulation of the bath
particles in front of the tracer which hinders fluctuations with $X_T(E,t)>%
\overline{X_T(E,t)}$. At the same time, the fluctuations with $X_T(E,t)<%
\overline{X_T(E,t)}$ are suppressed by the driving force. This leads to the
mean-square deviation $\overline{\sigma^{2}(t)}=(\overline{X_T(E,t)})^2-%
\overline{X_T^2(E,t)}$ growing only as $\sqrt{t}$ (see the numerical results
in Fig.2, where the $\sqrt{\overline{\sigma^2(t)}}$ is plotted versus $%
\sqrt{t}$), in a striking contrast to the linear time dependence for
conventional driven diffusive motion. Finally, within the same
''deterministic'' picture we obtain

\[
\hat{L}_{trac}(f_{\pm 1}) \; \approx \; \mp \; \overline{\dot{X}_{T}(E,t)}
\; (1 \; - \; f_{\pm 1}) \quad (20)
\]

Combining Eqs.(17) to (20) we obtain the following continuous-space equation

\[
\dot{f}_{\lambda} \; = \; \frac{1}{2} \; \frac{\partial^{2} f_{\lambda}}{%
\partial \lambda^{2}} \; + \; \overline{\dot{X}_{T}(E,t)} \; \frac{\partial
f_{\lambda}}{\partial \lambda}, \quad (21)
\]
which is to be solved subject to the boundary conditions
\[
\dot{f}_{\lambda =\pm 1}\;=\;\pm \;\left .\frac 12\;\frac{\partial f_\lambda
}{%
\partial \lambda }\right |_{\lambda =\pm 1} \;\mp \;\overline{\dot{X}_T(E,t)}%
\;(1\;-\;f_{\pm 1})\quad (22)
\]
and boundary conditions at $\lambda \rightarrow \pm \infty $, defined in
Eq.(16).

\subsection{Solutions for $\lambda \geq 1$.}

Let us consider first the behavior of $f_\lambda $ for $\lambda \geq 1$. We
introduce a scaled variable $\omega =(\lambda -1)/\overline{X_T(E,t)}$, ($%
0\leq \omega \leq \infty $). In terms of $\omega $ Eq.(21) takes the
following form
\[
\frac{d^{2} f_{\omega}}{d \omega^{2}} \; + \; \alpha \; (\omega + 1) \;
\frac{d f_{\omega}}{d \omega} \; = \; 0, \quad (23)
\]
in which we have set
\[
\alpha \; = \; \frac{d}{dt} \; (\overline{X_{T}(E,t)})^{2}, \quad (24)
\]
and which is to be solved subject to the boundary conditions
\[
f_{\omega = \infty} \; = \; C_{0}; \; \left . \frac{d f_{\omega}}{d
\omega}\right |_{\omega
= 0} \; = \; \alpha \; (1 \; - \; f_{\omega = 0}) \quad (25)
\]

We hasten to remark that such an approach presumes already that $\alpha $ is
a time-independent constant and thus $\overline{X_T(E,t)}$ grows as $\sqrt{%
\alpha t}$. However, such an assumption will be seen to be self-consistent
if it turns out to be possible (and we set out to show that this is actually
the case) to find the solution of Eq.(23) satisfying the boundary conditions
in Eqs.(25).

The appropriate solution to Eq.(23), which satisfies the boundary condition
at infinity, reads

\[
f_\omega \;=\;\frac{(C_0\;-\;f_{\omega =0})}{I_{+}}\int_0^\omega d\omega
\;\exp (-\alpha \;(\omega +\omega ^2/2))\;+\;f_{\omega =0},\quad (26)
\]
where
\[
I_{+}(\alpha )\equiv I_{+}\;=\;\int_0^\infty d\omega \;\exp (-\alpha
\;(\omega +\omega ^2/2))\;=\;\sqrt{\frac \pi {2\alpha }}\exp (\alpha
/2)\;[1\;-\;erf(\sqrt{\alpha /2})],\quad (27)
\]
$%
\mathop{\rm erf}
(x)$ being the error function. The value of $f_{\omega =0}$, which is the
probability of having the r-h-s adjacent to the tracer site vacant, is to be
defined from the boundary condition at $\omega =0$, Eq.(25), which gives

\[
\frac{C_0\;-\;f_{\omega =0}}{I_{+}}\;=\;\alpha \;(1\;-\;f_{\omega =0})\quad
(28)
\]
We note here that $\alpha $ is as yet unspecified parameter.

\subsection{Solutions for $\lambda \leq - 1$.}

Consider next the behavior of $f_\lambda $, Eq.(21), past the tracer particle,
i.e.
in the domain $\lambda
\leq -1$. Now we choose the scaled variable $\theta =-(\lambda +1)/\overline{%
X_T(E,t)}$, ($0\leq \theta \leq \infty $). In terms of $\theta $ Eq.(21)
takes the form

\[
\frac{d^2f_\theta }{d\theta ^2}\;+\;\alpha \;(\theta -1)\;\frac{df_\theta }{%
d\theta }\;=\;0,\quad (29)
\]
with boundary conditions

\[
f_{\theta = \infty} \; = \; C_{0}; \; \frac{d f_{\theta}}{d \theta}_{|\theta
= 0} \; = \; - \; \alpha \; (1 \; - \; f_{\theta = 0}) \quad (30)
\]

The solution of Eq.(29), which satisfies the boundary at $\theta =\infty $
reads

\[
f_{\theta} \; = \; \frac{(C_{0} \; - \; f_{\theta=0})}{I_{-}}
\int^{\theta}_{0} d\theta \; \exp(\alpha \; (\theta - \theta^{2}/2)) \; + \;
f_{\theta=0}, \quad (31)
\]
where

\[
I_{-}(\alpha )\equiv I_{-}\;=\;\int_0^\infty d\theta \;\exp (\alpha
\;(\theta -\theta ^2/2))\;=\;\sqrt{\frac \pi {2\alpha }}\exp (\alpha
/2)\;[1\;+\;erf(\sqrt{\alpha /2})] \quad (32)
\]
The constant of integration $f_{\theta =0}$ (which defines the probability
of having the l-h-s adjacent to the tracer site vacant) is to be determined
from the boundary condition at $\theta =0$, which gives

\[
\frac{C_{0} \; - \; f_{\theta=0}}{I_{-}} \; = \; - \; \alpha \; (1 \; - \;
f_{\theta=0}) \quad (33)
\]

Now we have a system of equations (28),(33) and (5),(24), which determine
the behavior of $\overline{X_T(E,t)}$, $f_{\omega =0}$ and $f_{\theta =0}$.
In the following subsection we, construct a closed-form transcendental
equation for the parameter $\alpha $.

\subsection{Closed-form equation for $\alpha$.}

We may formally rewrite Eq.(5) in terms of the parameters $f_{\omega=0}$ and
$f_{\theta=0}$ as

\[
\overline{\dot{X}_T(E,t)}\;=\;p\;f_{\omega =0}\;-\;q\;f_{\theta =0}\quad
(34)
\]
Further, using Eqs.(34),(24) and (28) we have

\[
\frac{(C_{0} \; - \; f_{\omega=0})}{2 I_{+} \overline{X_{T}(E,t)}} \; = \;
(p \; f_{\omega=0} \; - \; q \; f_{\theta=0}) (1 \; - \; f_{\omega=0} )
\quad(35)
\]

Let us note now that the r.h.s. of Eq.(35) tends to zero as $\overline{%
X_T(E,t)}$ grows. This means that (except for the case $C_0=1$, when Eq.(35)
has only a trivial solution $f_{\omega =0}=1$) the value $pf_{\omega =0}$
should tend to $qf_{\theta =0}$ as $t\rightarrow \infty $. Expressing $%
f_{\omega =0}$ as

\[
f_{\omega =0}\;=\;\frac{q\;f_{\theta =0}}p\;+\;\phi ,\quad (36)
\]
where $\phi $ is a deviation from the equilibrium value $f_{\omega =0}$, and
assuming that at sufficiently large times this deviation is small enough to
allow linearization of equation near the equilibrium, we have

\[
\phi \; \approx \; \frac{p \; C_{0} \; - \; q \; f_{\theta=0}}{2 \; p \;
I_{+} (p \; - \; q f_{\theta=0}) \; \overline{X_{T}(E,t)}} \quad (37)
\]

Now, from Eq.(33) we express $f_{\theta=0}$ through the parameter $\alpha$,
which gives

\[
f_{\theta=0} \; = \; \frac{C_{0} \; + \; \alpha \; I_{-}}{1 \; + \; \alpha
\; I_{-}}, \quad (38)
\]
and substituting Eq.(38) into the Eq.(37) we get

\[
\phi \; \approx \; \frac{C_{0} \; (p - q)\; + \; \alpha \; I_{-} (p \; C_{0}
\; - \; q)}{2 \; p \; I_{+} \; (p \; - \; q \; C_{0} \; + \; (p - q) \;
\alpha \; I_{-} ) \; \overline{X_{T}(E,t)}} \quad (39)
\]

Finally, noticing that $\overline{\dot{X}_T(E,t)}=p\phi $, and using the
definition of the parameter $\alpha $, Eq.(24), we obtain the following
closed transcendental equation for $\alpha $

\[
\alpha \; = \; \frac{C_{0} \; (p - q) \; + \; \alpha \; I_{-} \; (p \; C_{0}
\; - \; q)}{ I_{+} \; (p \; - \; q \; C_{0} \; + \; (p - q) \; \alpha \;
I_{-} )}, \quad (40)
\]
which, after some algebraic transformations, can be cast into a compact,
symmetric form

\[
(\alpha \;I_{+}(\alpha )\;+\;\frac{q-p\;C_0}{p-q})\;(\alpha \;I_{-}(\alpha
)\;+\;\frac{p-q\;C_0}{p-q})\;=\;\frac{p\;q\;(1-C_0)^2}{(p-q)^2}\quad (41)
\]
where the $\alpha $ dependence of $I_{\pm }$ (Eqs. (27) and (32)) has been
made explicit.

A simple analysis shows that Eq.(41) has positive bounded solutions $\alpha
(p,C_0)$ for any values of $p$, $q=1-p$ and $C_0$ (except the trivial case $%
C_0=1$ when bath particles are absent). This means, in turn, that our
approach to the solutions of coupled Eqs.(23),(25),(29) and (30) is
self-consistent and $\overline{X_T(E,t)}$ actually grows in proportion to $%
\sqrt{t}$. In the trivial case when $C_0=1$ (when one expects, of course,
the linear growth of $\overline{X_T(E,t)}$) Eq.(41) reduces to

\[
(\alpha \; I_{+} \; - \; 1) \; (\alpha \; I_{-} \; + \; 1) \; = \; 0 \quad
(42)
\]
whose root is given by

\[
\alpha I_{+} = 1, \quad (43)
\]
which means that $\alpha = \infty$ and thus $\overline{X_{T}(E,t)}$ grows at
a faster rate than $\sqrt{t}$.

Behavior of the prefactor $\alpha$ is presented in Fig.1, in which we plot
the numerical solution of Eq.(41) for different values of the transition
probabilities and different concentrations $C_{0}$. In the next subsection
we present analytical estimates of the behavior of $\alpha$ in several
limiting situations.

\subsection{Analytical estimates of limiting behaviors of the parameter $%
\alpha$.}

We start by considering situations in which the parameter $\alpha $ can be
expected to be small; namely, when $p$ is close to $q$ (in other words when $%
p$ is only slightly above $1/2$) or when the vacancy concentration is small,
$C_0\ll 1$. In the limit of small $\alpha $ the leading terms in Eqs.(27)
and (32) behave as

\[
I_{\pm} \; \approx \; \sqrt{\frac{\pi }{2 \alpha}} \quad (44)
\]
and Eq.(41) takes the form

\[
( \sqrt{\frac{\pi \alpha}{2}} \; + \; \frac{q - p \; C_{0}}{p - q}) \; (
\sqrt{\frac{\pi \alpha}{2}} \; + \; \frac{p - q \; C_{0}}{p - q}) \; = \;
\frac{p \; q \; (1 - C_{0})^{2}}{(p - q)^{2}}, \quad (45)
\]
and gives for small $\alpha$, $\alpha \ll 1$,

\[
\alpha \; \approx \; \frac{2}{\pi} (\frac{C_{0} \; (p - q)}{1 - C_{0}})^{2}
\quad (46)
\]
Consequently, in this limit the explicit formula for the mean displacement
will read

\[
\overline{X_T(E,t)}\;\approx \;\frac{C_0\;(p-q)}{1-C_0}\;\sqrt{\frac{2\;t}\pi
}\;=\;\frac{C_0\;\tanh (\beta E/2)}{1-C_0}\sqrt{\frac{2\;t}\pi }\quad (47)
\]

At this point it might be instructive to recall the analysis in \cite{fer}
which concerned the validity of the Einstein relation for one-dimensional
hard-core lattice gases. Define the ''diffusion coefficient'' of the tracer
particle in the symmetric case as $D=lim_{t\to \infty }D(t)$, where $D(t)=%
\overline{X_T^2(E=0,t)}/(2t)$ and the ''mobility'' as $\mu =lim_{E\to
0}(U(E)/E)$, where the stationary velocity is given by $U(E)=lim_{t\to
\infty }U(E,t)$, $U(E,t)=X_T(E,t)/t$. In \cite{fer} it was shown that for
the infinitely large system the Einstein relation, i.e. the equation $\mu
=\beta D$, holds trivially, since both $D$ and $\mu $ are equal to zero. A
somewhat stronger result has been obtained for the case when the
one-dimensional lattice is a ring of length $l$. It was shown that here both
$D$ and $\mu $ are finite, vanish with the length of the ring as $1/l$ and
obey the Einstein relation $\mu (l)=\beta D(l)$. Our result in Eq.(47)
reveals that for the infinite lattice Einstein relation is valid for the
time-dependent ''diffusion coefficient'' and mobility, i.e. equation $\mu
(t)\;(=lim_{E\to 0}(U(E,t)/E))\;=\;\beta D(t)$ holds exactly, provided that
the time is sufficiently large such that both Eqs.(1) and (47) are valid.

Consider next the situation when $\alpha$ is expected to be large, i.e. when
$C_{0}$ is close to unity. Expanding the functions $I_{+}$ and $I_{-}$ as

\[
I_{+} \; \approx \; \frac{1}{\alpha} \; - \; \frac{1}{\alpha^{2}} \;
\rightarrow \; 0, \; \text{when} \; \alpha \; \rightarrow \; \infty, \quad
(48)
\]
and
\[
I_{-} \; \approx \; \sqrt{\frac{2 \pi}{\alpha}} \exp(\alpha/2) \;
\rightarrow \; \infty, \; \text{when} \; \alpha \; \rightarrow \; \infty,
\quad (49)
\]
we arrive at the following equation

\[
(1\;+\;\frac{q-p\;C_0}{p-q}\;-\;\frac 1\alpha )\;\;=\;\frac{p\;q\;(1-C_0)^2}{%
(p-q)^2}\;(\sqrt{2\pi \alpha }\;\exp (\alpha /2)\;+\;\frac{p-q\;C_0}{p-q}%
)^{-1}\quad (50)
\]
One may readily notice that when $\alpha $ is large, the r.h.s. of Eq.(50)
tends to zero and thus we have

\[
\alpha \;\approx \;\frac{p-q}{p\;(1-C_0)}\quad (51)
\]
Eq.(51) allows formulation of the conditions when $\alpha $ is large more
precisely: this happens when $C_0$ obeys the inequality $1>C_0\gg q/p=\exp
(-\beta E)$.

In this limit the mean displacement of the tracer particle obeys

\[
\overline{X_{T}(E,t)} \; \approx \; \sqrt{\frac{(p - q) \; t}{p \; ( 1 -
C_{0})}} \; = \; \sqrt{\frac{(1 - \exp(- \beta E)) \; t}{1 - C_{0}}} \quad
(52)
\]

Finally, let us consider the case of the directed walk in a one-dimensional
lattice gas, when $p = 1$ and $q = 0$, which was previously examined in \cite
{burd}. In this case Eq.(41) simplifies considerably and reads

\[
\alpha_{\infty} I_{+} \; = \; C_{0} \quad (53)
\]
When $C_{0}$ is small, $C_{0} \ll 1$, we have from Eq.(53)

\[
\alpha_{\infty} \; \approx \; \frac{2 C_{0}^{2}}{\pi}, \quad (54)
\]
which is consistent with the result in Eq.(47), while in the limit $C_{0}
\approx 1$, Eq.(53) yields

\[
\alpha_{\infty} \; \approx \; \frac{1}{1 - C_{0}}, \quad (55)
\]
i.e. exactly the behavior in Eq.(51) with $p = 1$ and $q = 0$.

\section{Numerical simulations.}

In order to check the analytical predictions of Sections IV and V Monte
Carlo (MC) simulations
have been performed. The simulation algorithm was defined as follows: The total
number of particles in the system was held constant at $399$ ($398$ bath
particles and one tracer). We constructed a one-dimensional regular lattice
of unit spacing and length $2L+1$, sites of which were labelled by integers
on  the interval $[-L,L]$. For each $C_p,$ the length was chosen as $L=%
\frac{399}{C_p}.$ The number of particles and, thus, $L$ was chosen to be
large enough so that over the time scale of the simulations the mean vacancy
density near $L_{\pm }$ was unaffected by the dynamics of the tracer which
always occupied the lattice site $0$ at $t=0.$   At the zero moment of MC
time particles were placed randomly  on the
lattice with the prescribed mean concentration and the constraint that no two
particles can simultaneously occupy the same site. The tracer particle was
placed at the origin. The subsequent particle dynamics employed in simulations
follows the definitions of Section 2 closely. At each MC time
step we select, at random, one particle and let it choose a potential jump
direction. If the selected particle is a bath particle, it chooses the
direction of the jump - to the right or to the left, with the probability $%
1/2$. If the selected particle is the tracer - it chooses the hop to the
right with probability $p$ and the hop to the left with probability $q$, $%
q=1-p$. For any particle the jump is instantly fulfilled if at this moment
of time the adjacent site in the chosen direction is vacant.

In simulations we followed the time evolution of the displacement of the
tracer particle and plotted it versus the  ''physical time'', which is the
time needed for each particle to move once, on average. Fig.2 presents
typical
behavior of the  average displacement calculated for fixed $q$
and $C_0$ (the actual parameter for the presented results are $q=0$ and $%
C_0=1/3,$ the results for other values of $C_0$ and $q$ are similar) plotted
versus $\sqrt{t}$. The dotted line in this figure shows the evolution of the
deviation from the mean average displacement $\sqrt{\overline{\sigma ^2}}$
which shows growth in proportion to $t^{1/4}$.

Numerical results for the time evolution of the mean displacement, performed
for different values of parameters $C_0$ and $p$, were used to obtain a
numerical evaluation of  the prefactor in Eq.(3) as a function of $C_0$ and
$p$. In Fig.3 we plot, for the particular case of totally directed walk
$q=0$,
the prefactor $\alpha $ versus the inverse concentration of vacancies. The
solid line represents the results of our analytic calculations while the
squares show the numerical results. In Fig.4, for fixed concentration of
the vacancies $C_0=1/3$ we present the comparison between the analytical
predictions of the prefactor $\alpha$ and the numerical  MC results.

\section{Conclusions.}

To conclude, we have examined the behavior of the mean displacement of a
driven tracer particle moving in a symmetric hard-core lattice gas. We have
shown that the mean displacement grows in proportion to $\sqrt{t}$, i.e. the
hard-core interactions hinder the ballistic motion of the tracer and
introduce effective frictional forces. The prefactors in the growth law
are determined implicitly, in a form of the transcendental equation, for
arbitrary magnitude of the driving force and arbitrary concentration of the
lattice gas particles. In several asymptotic limits we find explicit
formulae for these prefactors. Our analytical findings are in a very good
agreement with the results of numerical simulations.

\begin{center}
Acknowledgments.
\end{center}

The authors address special thanks to J.L.Lebowitz for many encouraging
discussions and interest in this work. We also acknowledge helpful
discussions with K.W.Kehr, E.G.D.Cohen and H.van Beijeren. S.F.B. and W.P.R.
acknowledge the  support of ONR Grant N 00014-94-1-0647 and, and S.F.B.  by
the University of Paris VI. G.O. acknowledges financial support from the
CNRS.

\begin{figure}
\caption{Theoretical values of the prefactor in the dependence of
the average tracer displacement $X(t)$ on $t^{1/2}$
vs the probability of the "back step", $q$.}
\label{Fig1}
\caption{Typical dependence of squared mean
tracer displacement and mean squared deviation on time.}
\label{Fig2}
\caption{Theoretical and experimental values of the prefactor in the dependence
of
the average tracer displacement, $X(t)$, on $t^{1/2}$ vs the inverse
concentration
of bath particles ($q=0$).}
\label{Fig3}
\caption{Theoretical and experimental values of the prefactor in the dependence
of
the average tracer displacement, $X(t)$, on $t^{1/2}$ vs $q$ ($C=1/3$).}
\label{Fig4}
\end{figure}

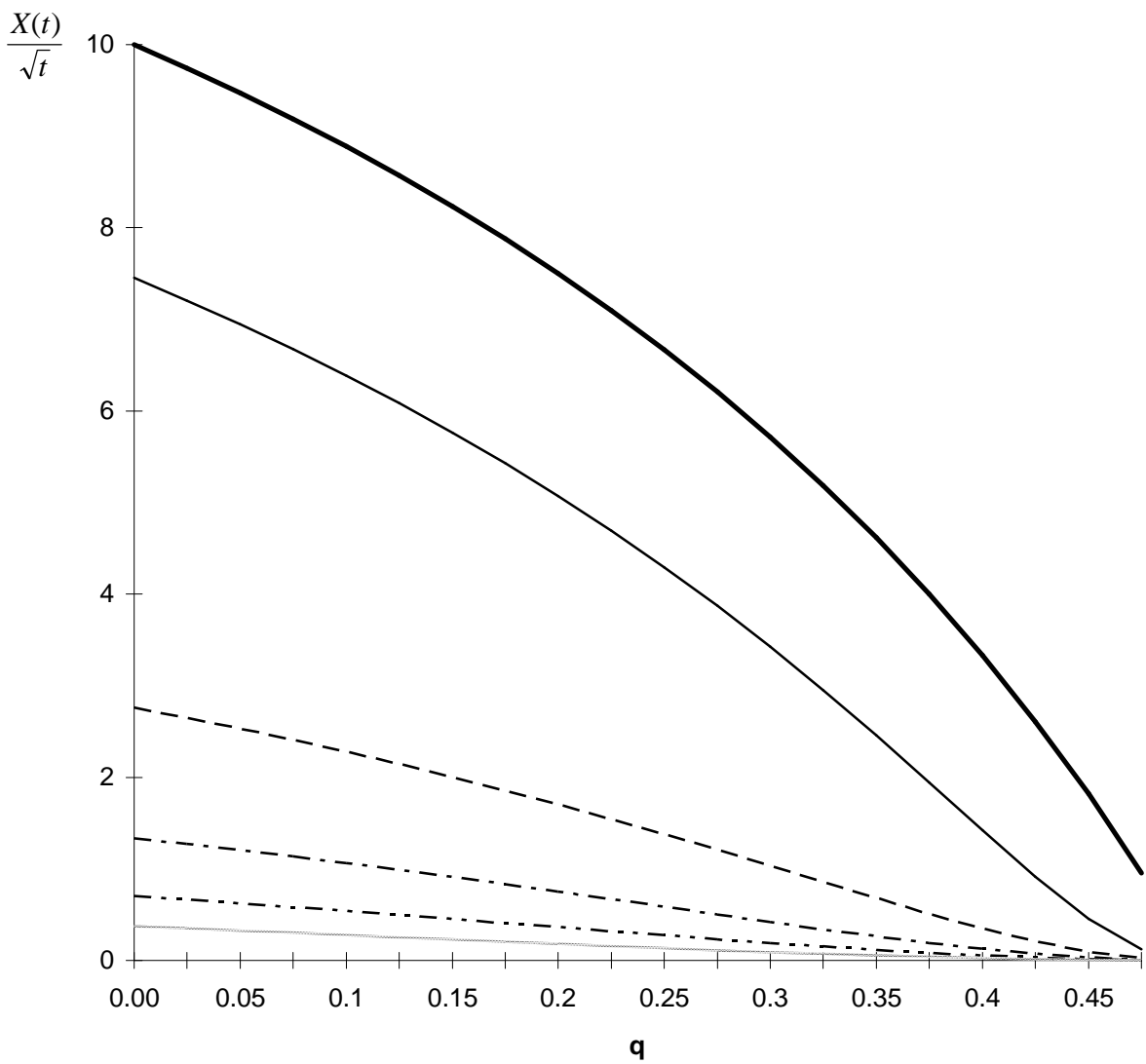
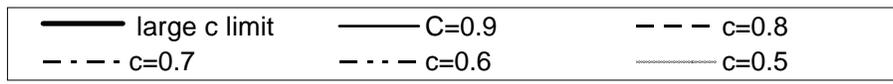
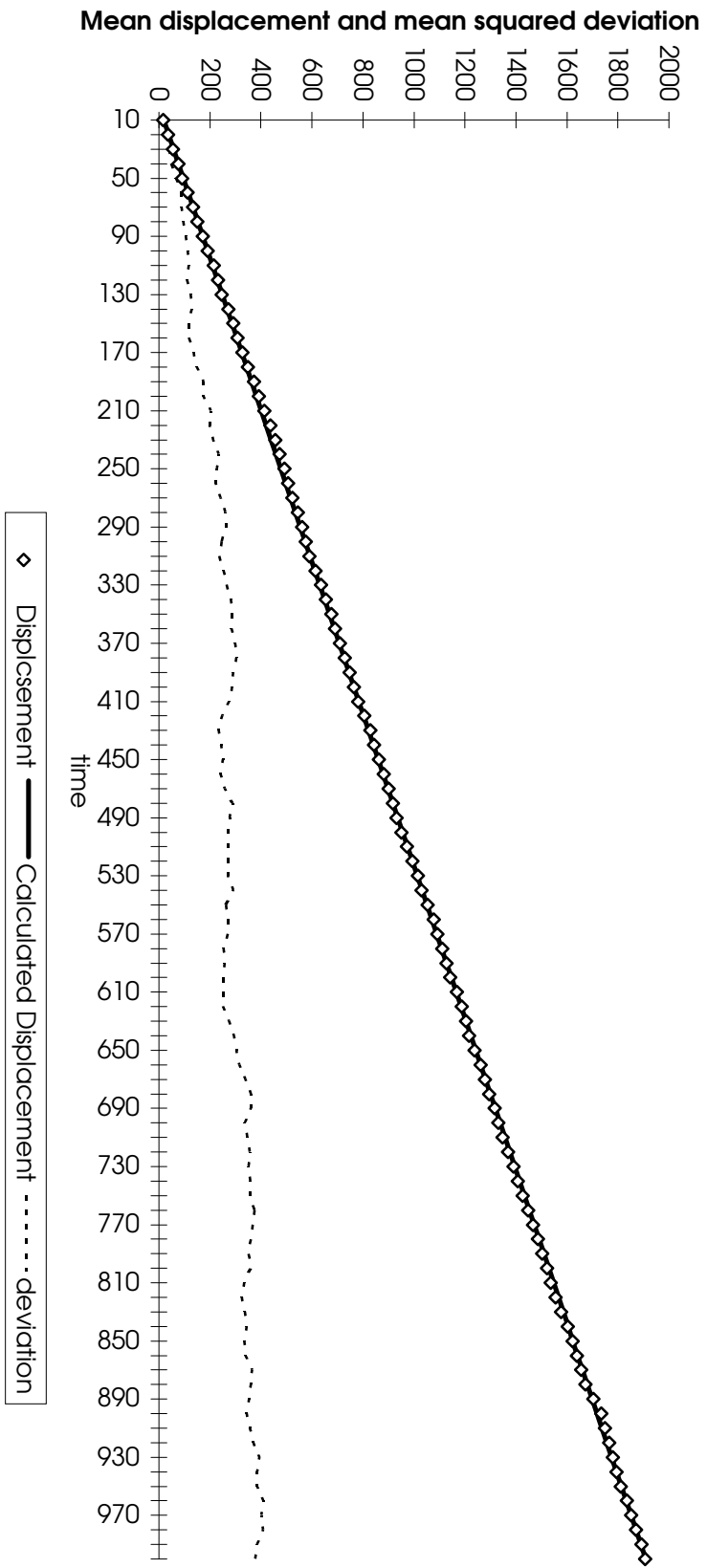
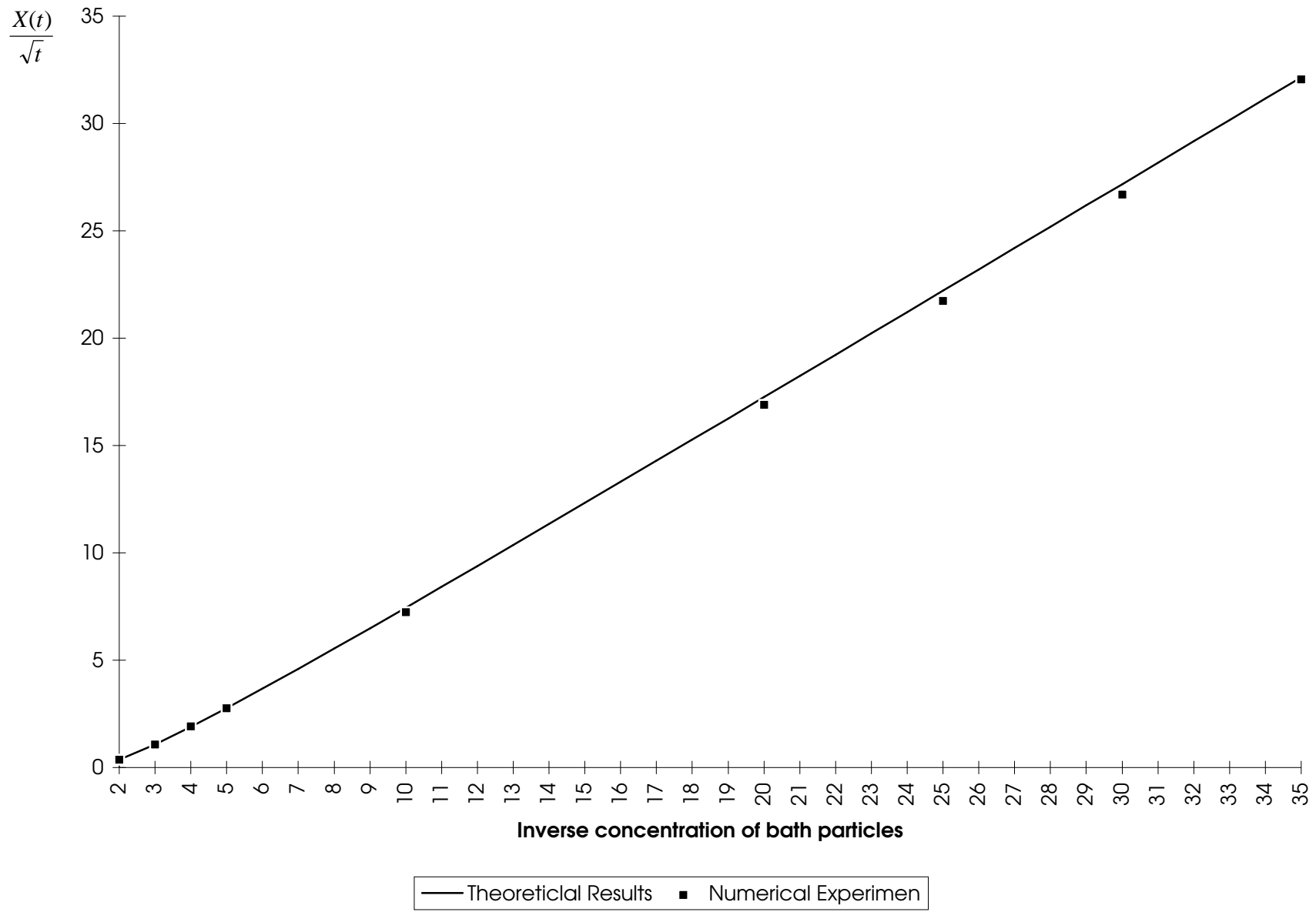
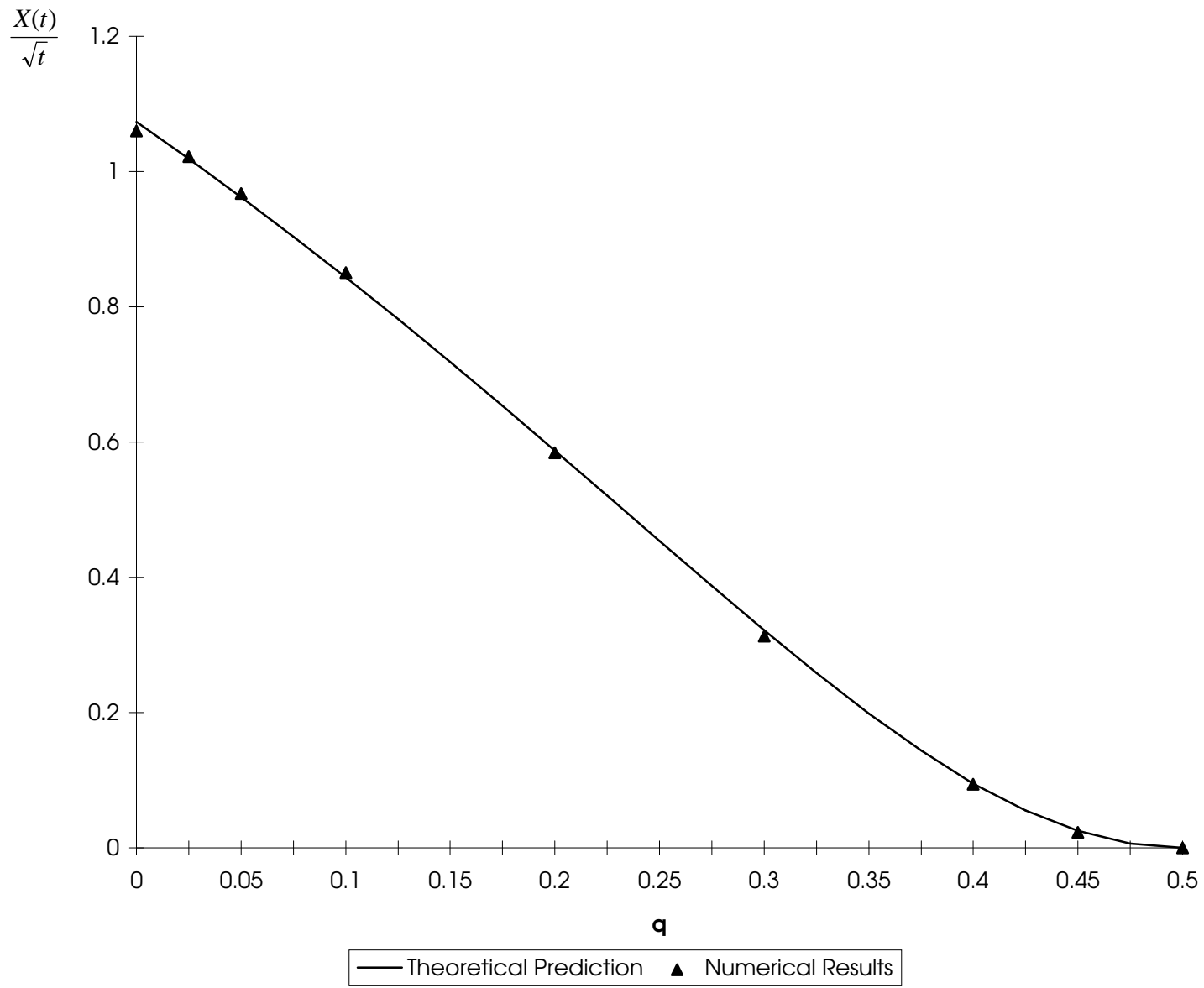
\end{document}